\documentclass[conference]{IEEEtran}
\IEEEoverridecommandlockouts
\usepackage{cite}
\usepackage{amsmath,amssymb,amsfonts}
\usepackage{algorithmic}
\usepackage{graphicx}
\usepackage{svg}
\usepackage{textcomp}
\usepackage{subfig}
\usepackage{hyperref}
\usepackage{xcolor}
\usepackage{hhline}
\usepackage{float}
\usepackage{enumitem}
\usepackage[utf8]{inputenc}
\def\BibTeX{{\rm B\kern-.05em{\sc i\kern-.025em b}\kern-.08em
    T\kern-.1667em\lower.7ex\hbox{E}\kern-.125emX}}
\begin{document}

\title{A Benchmark to Evaluate InfiniBand Solutions for Java Applications}

\author{
{\rm Stefan~Nothaas}\\
stefan.nothaas@hhu.de
\and
{\rm Fabian~Ruhland}\\
fabian.ruhland@hhu.de
\and
{\rm Michael~Schöttner}\\
michael.schoettner@hhu.de
\and
Heinrich-Heine-Universität Düsseldorf
} 

\author{\IEEEauthorblockN{Stefan~Nothaas}
\IEEEauthorblockA{\textit{Department of CS Operating Systems} \\
\textit{Heinrich-Heine-Universität}\\
Duesseldorf, Germany \\
stefan.nothaas@hhu.de}
\and
\IEEEauthorblockN{Fabian~Ruhland}
\IEEEauthorblockA{\textit{Department of CS Operating Systems} \\
\textit{Heinrich-Heine-Universität}\\
Duesseldorf, Germany \\
fabian.ruhland@hhu.de}
\and
\IEEEauthorblockN{Michael~Schoettner}
\IEEEauthorblockA{\textit{Department of CS Operating Systems} \\
\textit{Heinrich-Heine-Universität}\\
Duesseldorf, Germany \\
michael.schoettner@hhu.de}
}

\maketitle

\begin{abstract}
Low-latency network interconnects, such as InfiniBand, are commonly used in HPC centers and are even accessible with todays cloud providers offering equipped instances for rent. Most big data applications and frameworks are written in Java. But, the JVM environment alone does not provide interfaces to directly utilize InfiniBand networks.

In this paper, we present the ``Java InfiniBand Benchmark'' to evaluate the currently available (and supported) ``low-level'' solutions to utilize InfiniBand in Java. It evaluates socket- and verbs-based libraries using typical network microbenchmarks regarding throughput and latency. Furthermore, we present evaluation results of the solutions on two hardware configurations with 56 Gbit/s and 100 Gbit/s InfiniBand NICs. With transparency often traded for performance and vice versa, the benchmark helps developers with studying the pros and cons of each solution and support them in their decision which solution is more suitable for their existing or new use-case.
\end{abstract}


\section{Introduction}
\label{introduction}

RDMA capable devices have been providing high throughput and low-latency to HPC applications for several years \cite{top500}. With todays cloud providers offering instances equipped with InfiniBand for rent, such hardware is available to a wider range of users without the high costs of buying and maintaining it \cite{HASHEM201598}. Many application domains such as social networks \cite{facebook2, Liu:2016:ECI:2964797.2964815, Twitter}, search engines \cite{pagerank, Gulli:2005:IWM:1062745.1062789}, simulations \cite{doi:10.1093/bioinformatics/btt055} or online data analytics \cite{Desikan:2005:IPR:1062745.1062885, 6547630, DOI:10.1007/978-3-319-55699-4_20} require large processing frameworks and backend storages. Many of these are written in Java, e.g. big data batch processing frameworks \cite{hadoopecosystem, Kreps:2011aa}, databases \cite{cassandra, Ignite} or backend storages/caches \cite{hazelcast, gemfire, infinispan, oraclecoherance}. 

These applications benefit from the rich environment Java offers including automatic garbage collection and multi-threading utilities. But, the choices for inter-node communication on distributed applications are limited to Ethernet-based socket-interfaces (standard ServerSocket or NIO) on the commonly used JVMs OpenJDK and Oracle. They do not provide support for low-latency InfiniBand hardware. But, there are external solutions available each with pros and cons. 

This raises questions if a developer wants to chose a suitable solution for a new use-case or an existing application: What's the throughput/latency on small/large payload sizes? Is the performance sufficient when trading it for transparency requiring less to no changes to the existing code? Is it worth considering developing a custom solution based on the native API to gain maximum control with chances to harvest the performance available by the hardware?

In this paper, we address these questions by presenting a ``Java InfiniBand (JIB) benchmark'' to evaluate existing solutions to leverage the performance of InfiniBand hardware in Java applications. The modular benchmark currently provides implementations to evaluate three socket-based libraries and implementations, IP over InfiniBand, libvma and JSOR, as well as two verbs-based implementations, native C-verbs and jVerbs. This paper focuses on the fundamental performance metrics of low-level interfaces and \textit{not} on higher-level network subsystems with connection management, complex pipelines and messaging primitives, e.g. MPI. We discuss and evaluate these in a separate publication \cite{DBLP:journals/corr/abs-1812-01963}. We used our benchmark to evaluate the listed solutions on two hardware configurations with 56 Gbit/s and 100 Gbit/s InfiniBand NICs. The contributions of this paper are:

\begin{itemize}[noitemsep,topsep=0pt,parsep=0pt,partopsep=0pt]
 \item An overview of existing Java InfiniBand solutions
 \item An extensible and open source benchmark to easily evaluate solutions to use InfiniBand in Java applications
 \item Extensive evaluation of existing Java libraries with 56 Gbit/s and 100 Gbit/s hardware
\end{itemize}

The remaining paper is structured as follows: Section \ref{related_work} discusses related work with socket-based (\S \ref{related_work_socket}) and verbs-based (\S \ref{related_work_verbs}) libraries. Section \ref{jib} presents the JIB Benchmark Suite which is used to evaluate two verbs-based solutions and three socket-based solutions in the following Section \ref{evaluation} regarding overhead (\S \ref{eval_overhead}), uni-directional (\S \ref{eval_uni}) and bi-directional (\S \ref{eval_bi}) throughput, as well as one-sided latency (\S \ref{eval_lat}) and full round-trip-time using a ping-pong benchmark (\S \ref{eval_pingpong}). Conclusions are presented in Section \ref{conclusions}.

\section{Related Work}
\label{related_work}

This section elaborates on existing ``low-level'' solutions/libraries that can be used to leverage the performance of InfiniBand hardware in Java applications. This does not include network or messaging stacks/subsystems implementing higher-level primitives such as the Massage Passing Interface, e.g. Java-based FastMPJ \cite{Exposito:2014aa} providing a special transport to use InfiniBand hardware. To the best of our knowledge, there is no benchmark available to evaluate InfiniBand solutions in Java.

\subsection{Socket-based Libraries}
\label{related_work_socket}

The socket-based libraries redirect the send and receive traffic of socket-based applications transparently over InfiniBand host channel adapters (HCAs) with or without kernel bypass depending on the implementation. Thus, existing applications do not have to be altered to benefit from improved performance due to the lower latency hardware compared to commonly used Gigabit Ethernet. The following three libraries are still supported to date and evaluated in Section \ref{evaluation}.

\textbf{IP over InfiniBand (IPoIB)} \cite{ipoib} is not a library but actually a kernel driver that exposes the InfiniBand device as a standard network interface (e.g. \textit{ib0}) to the user space. Socket-based applications do not have to be modified but use the specific interface. However, the driver uses the kernel's network stack which requires context switching (kernel to user space) and CPU resources when handling data. Naturally, this solution trades performance for transparency.

\textbf{libvma} \cite{libvma} is a library developed by Mellanox and included in their OFED software package \cite{mellanox}. It is pre-loaded to any socket-based application (using \textit{LD\_PRELOAD}). It enables full bypass of the kernel network-stack by redirecting all socket-traffic over InfiniBand using unreliable datagram with native C-verbs. Again, the existing application code does not have to be modified to benefit from increased performance.

\textbf{Java Sockets over RDMA (JSOR)} \cite{jsor} redirects all socket-based data traffic in Java applications using native verbs, similar to libvma. It uses two paths for implementing transparent socket streams over RDMA devices. The "fast data path" uses native verbs to send and receive data and the "slow control path" manages RDMA connections. JSOR is developed by IBM on only available in their proprietary J9 JVM.

The following libraries are also known in literature but are not supported or maintained anymore.

The \textbf{Sockets Direct Protocol (SDP)} \cite{Goldenberg2006ArchitectureAI} redirects all socket-based traffic of Java applications over RDMA with kernel-bypass. It supported all available JDKs since Java 7 and was part of the OFED package until it was removed with OFED version 3.5 \cite{ofed35releasenotes}.

\textbf{Java Fast Sockets (JFS)} \cite{Taboada:2008:JFS:1456731.1457122} is an optimized Java socket implementation for high speed interconnects. It avoids serialization of primitive data arrays and reduces buffering and buffer copying with shared memory communication as its main focus. However, JFS relies on SDP (deprecated) for using InfiniBand hardware.

\textbf{Speedus} \cite{speedus} is a native library that optimizes data transfers for applications especially on intra-host and inter-container communication by bypassing the kernel's network stack. It is also advertised to support low-latency networking hardware for inter-node communication. But, the latest available version to date (2016-09-08) does not include such support.

\subsection{Verb-based Libraries}
\label{related_work_verbs}

Verbs are an abstract and low-level description of functionality for RDMA devices (e.g. InfiniBand) and how to program them. Verbs define the control and data paths including RDMA operations (write/read) as well as messaging (send/receive). RDMA operations allow reading or writing directly from/to the memory of the remote host without involving the CPU of the remote. Messaging follows a more traditional approach by providing a buffer with data to send and the remote providing a buffer to receive the data to. 

The programming model differs heavily from traditional socket-based programming. Using different types of asynchronous queues (send, receive, completion) as communication endpoints. The application uses different types of work-requests for sending and receiving data. When handling data to transfer, all communication with the HCA is executed using these queues. The following libraries are verbs implementations that allow the user to program the RDMA capable hardware directly. The first two libraries presented are evaluated in Section \ref{evaluation}.

\textbf{C-verbs} are the native verbs implementation included in the OFED package \cite{ofed}. Using the Java Native Interface (JNI) \cite{Liang:1999:JNI:520155}, this library can be utilized in Java applications as well in order to create a custom network subsystem \cite{Exposito:2014aa} \cite{DBLP:journals/corr/abs-1812-01963}. Using the Unsafe class \cite{javaunsafe} or Java DirectByteBuffers, memory can be allocated off-heap to use it for sending and receiving data with InfiniBand hardware (buffers must be registered with a protection domain which pins the physical memory).

\textbf{jVerbs} \cite{Stuedi:2013:JUL:2523616.2523631} are a proprietary verbs implementation for Java developed by IBM for their J9 JVM. Using a JNI layer, the OFED C-verbs implementation is accessed. ``Stateful verb methods'' (\textit{StatefulVerbsMethod} Java objects) encapsulate the verb to call including all parameters with parameter serialization to native space. Once the object is prepared, it can be executed which actually calls the native verb. These objects can be re-used for further calls with the same parameters to avoid repeated serialization to native space and creating new objects which would burden garbage collection.

\textbf{Jdib} \cite{jdib} is a library wrapping native C-verbs function calls and exposing them to Java using a JNI layer. According to the authors, various methods, e.g. queue pair data exchange on connection setup, are abstracted to create an easier to use API for Java programmers. The fundamental operations to create protection domains, create and setup queue pairs, as well as posting data-to-send to queues and polling the completion queue seem to wrap the native verbs and do not introduce additional mechanisms like jVerbs's stateful verb calls. We were not able to obtain a copy of the library for evaluation.

\section{A Benchmark for Evaluating InfiniBand Libraries for Java}
\label{jib}

To evaluate and compare the different libraries available, a common set of benchmarks had to be implemented for two programming languages (C and Java) and two programming models (sockets and verbs). Existing solutions like the iperf \cite{iperf} tools for TCP/UDP or the ibperf tools included in the OFED package \cite{ofed} do not cover all libraries we want to evaluate and do not implement all necessary benchmark types.

In this paper, we want to evaluate most of the available and still maintained libraries (\S \ref{related_work}) in a fundamental point-to-point setup regarding throughput and latency. Like other benchmarks (e.g. OSU \cite{osu}), we want to determine the maximum throughput on uni-directional and bi-directional communication (e.g. application pattern asynchronous ``messaging''), as well as one-sided latency and full round-trip-time (RTT) with a pingpong communication pattern (e.g. application pattern ``request-response''). These benchmarks allow us to determine the fundamental performance of the presented solutions and are commonly used to evaluate network hardware or applications \cite{iperf, ofed, osu}. The evaluation of higher-level primitives, e.g. collectives, and all-to-all communication is not possible with fundamental low-level interfaces. These require a higher-level networking stack with connection management and a complex pipeline which is not the focus of this paper.

The Java InfiniBand Benchmark (JIB) provides implementations of the listed benchmarks for two verbs-based solutions (C-verbs, jVerbs) and three socket-based solutions (IPoIB, libvma, JSOR). It is open source and available at Github \cite{githubjib}. Each benchmark run is configurable using command line parameters such as the benchmark type (uni-/ bi-directional, one-sided latency or ping-pong), the message size to send/receive and the number of messages to send/receive. For convenience, we refer to the payload size sent as messages independent of how it is transferred (e.g. sockets, verbs RDMA or verbs messaging). The context and all buffers are initialized before the benchmark is started. Afterwards, the current instance connects to the remote specified via command line parameters. Once the connection is established, a dedicated thread is started for sending data and another thread for receiving. Today, we can expect this to run on a multicore system with at least two physical cores to ensure that the send and receive thread can be run simultaneously to avoid blocking one another. The benchmark instance sends the specified number of messages to the remote and measures the time it takes to send the messages. Furthermore, we utilize the performance counters of the InfiniBand HCA to determine the overhead added by any software defined protocol which is especially relevant for the socket-based libraries (\S \ref{eval_overhead}).

For socket-based libraries, the benchmark is implemented in Java using TCP sockets with the ServerSocket, DataInputStream and DataOutputStream classes. Sending and receiving data is executed synchronously in a single loop on each thread. No further adjustments are required because all libraries redirect the normal send and receive calls of the socket libraries. With IPoIB, we use the address of the exposed \textit{ib0} device. The libvma library is pre-loaded to the benchmark using \textit{LD\_PRELOAD}. In order to use JSOR, we run the benchmark in the J9-JVM and provide a configuration file specifying IP-addresses whose traffic is redirected over the RDMA device.

The verbs-based benchmarks are implemented in C and Java. Both implementations use RC queue pairs for RDMA and message operations. UD queue pairs can also be used for message operations but this option is currently not implemented. On RDMA operations, we did not include immediate data with a work request which would require a work completion on the remote (optional for signalling incoming RDMA operations on the remote). When sending RDMA operations to the HCA to determine the maximum throughput, we do not repeatedly add one work request to the send queue, then poll the completion queue waiting for that single work request to complete. This pattern is commonly used in examples \cite{rdmamojo} and even larger applications \cite{ramcloudgithub} but does not yield optimal throughput because the queue of the HCA runs empty very frequently. To keep the HCA busy, the send queue must be kept filled at all times. Thus, we fill up the send queue to the maximum size configured, first. Then, we poll the completion queue and once at least one completion is available and processed, we immediately fill the send queue again. Naturally, this pattern cannot be applied to the ping-pong latency benchmark executing a request-response pattern.

This data path is implemented in both, the C-verbs and jVerbs implementation. The C implementation uses the verbs implementation included in the OFED package and serves as a reference for comparing the maximum possible performance. To establish a remote connection, queue pair information is exchanged using a TCP socket. The jVerbs implementation has to implement the operations of the data path using the previously described stateful verbs methods. Thus, the sending of data on the throughput benchmark had to be altered slightly. A single stateful verb call for posting work requests to the send queue always posts 10 elements. Hence, new work requests are added to the send queue once at least 10 work completions were polled from the completion queue. We create all stateful verbs calls before the benchmark and re-use them to avoid performance penalties. On connection creation, queue pair information is exchanged with the API provided by jVerbs which is using the RDMA connection manager.
    
\section{Evaluation}
\label{evaluation}

In this Section, we present the results of the evaluation of the socket-based libraries/implementations \textbf{IPoIB}, \textbf{libvma} and \textbf{JSOR} and the verbs-based libraries \textbf{C-verbs} and \textbf{jVerbs} using our benchmark suite (\S \ref{jib}). We analyze and discuss basic performance metrics regarding throughput and latency using typical benchmarks with a two node setup with 56 Gbit/s and 100 Gbit/s interconnects. A summary of the benchmarks executed with each library/implementation is given in Table \ref{eval_table}. Due to space constraints, we limit the discussion of the results to selected conspicuities of the plotted data.

\begin{table}[H]
\centering
\begin{tabular}{|l|c|c|c|c|c|}
\hline
\textbf{Library/Benchmark} & \multicolumn{1}{l|}{\textbf{OV}} & \multicolumn{1}{l|}{\textbf{Uni-dir}} & \multicolumn{1}{l|}{\textbf{Bi-dir}} & \multicolumn{1}{l|}{\textbf{Lat}} & \multicolumn{1}{l|}{\textbf{PingPong}} \\ \hline
C-verbs rdmaw &                               & x                            & x                           & x                            &                               \\ \hline
C-verbs rdmar &                               & x                            & x                           & x                            &                               \\ \hline
C-verbs msg   & x                             & x                            & x                           & x                            & x                             \\ \hline
jVerbs rdmaw  &                               & x                            & x                           & x                            &                               \\ \hline
jVerbs rdmar  &                               & x                            & x                           & x                            &                               \\ \hline
jVerbs msg    & x                             & x                            & x                           & x                            & x                             \\ \hhline{|=|=|=|=|=|=|}
IPoIB         & x                             & x                            & x                           & x                            & x                             \\ \hline
JSOR          & x                             & x                            & error                       & x                            & x                             \\ \hline
libvma        & x                             & x                            & x                           & x                            & x                             \\ \hline
\end{tabular}
\caption{Overview of libraries evaluated with benchmarks available. Abbreviations: OV = Overhead, rdmaw = RDMA write, rdmar = RDMA read, msg = messaging verbs} 
\label{eval_table}
\end{table}

We use the term ``message'' (msg) to refer to the unit of transfer which is equivalent to the data payload. The size of a message does refer to the payload size only and does not include any additional protocol or network layer overhead. Each throughput focused benchmark run transfers 100 million messages and each latency focused benchmark run transfers 10 million messages. Starting with 8 kB message size, the amount of messages is incrementally halved to avoid unnecessary long running benchmark runs. We evaluated payload sizes of 1 byte to 1 MB in power-of-two-increments. When discussing the results, we focus on the message rate on small messages with payload sizes less than 1 kB and on the throughput on middle sized and large messages starting at 1 kB. 

The throughput results are depicted as line plots with the left y-axis describing the throughput in million messages per second (mmps) and the right y-axis describing the throughput in MB/s. For the latency results, the left y-axis describes the latency in \textmu s and the right y-axis the throughput in mmps. The dotted lines always represent the message throughput while the solid lines represent either the throughput in MB/s or the latency in \textmu s, depending on the benchmark. For the overhead results, a single y-axis describes the overhead in percentage in relation to the amount of payload transferred on a logarithmic scale. On all plot types, the x-axis depicts the size of the payload in power-of-two increments from 1 byte to 1 MB. Each benchmark run was executed three times and the min and max as well as average of the three runs are visualized using error bars.

The following releases of software were used for compiling and running the benchmarks: Java 1.8, OFED 4.4-2.0.7, libvma 8.7.5, IBM J9 VM version 2.9, gcc 8.1.0. We ran our experiments on the following two setups with two nodes each:
\begin{enumerate}
 \item Mellanox ConnectX-3 HCA, 56 Gbit/s InfiniBand, MTU size 4096, Bullx blade with Dual socket Intel Xeon E5-2697v2 (2.7 GHz) 12 core CPUs, 128 GB RAM, CentOS 7.4 with the Linux Kernel version 3.10.0-693, SBE-820C with built-in switch 
 \item Mellanox ConnectX-5 HCA, 100 Gbit/s InfiniBand, MTU size 4096, Supermicro blade with Dual socket Intel Xeon Gold 6136 (3.0 GHz) 12 core CPUs, 128 GB RAM, CentOS 7.4 with the Linux Kernel version 3.10.0-693, Super Micro EDR-36 Chassis with built-in switch
\end{enumerate}

Flow steering must be activated for libvma to redirect all traffic over InfiniBand by setting the parameter \textit{log\_num\_mgm\_entry\_size} to \textit{-1} in the configuration file \textit{/etc/modprobe.d/mlnx.conf} for the InfiniBand kernel module. Otherwise, libvma falls back to sockets over Ethernet.

In the following subsections, we focus the analysis and discussion on the results with 100 Gbit/s hardware. Selected Figures depicting the results with 56 Gbit/s hardware are also included if they provide additional insights and value for discussion and comparision. Due to automated execution of the benchmarks, the naming in the figures differs slightly, e.g. ''JSocketBench(msg)`` refers to IPoIB. The other names are self-explanatory.

\subsection{Overhead}
\label{eval_overhead}

In this Section, we present the results of the overhead measurements of the described libraries/implementations. As overhead, we consider the additional amount of data that is sent along with the payload data of the user. This includes any data of any network layer down to the HCA. We measured the amount of data emitted by the port using the performance counter \textit{port\_xmit\_data} of the HCA.

\begin{figure}
	\centering
	\includegraphics[width=3.3in]{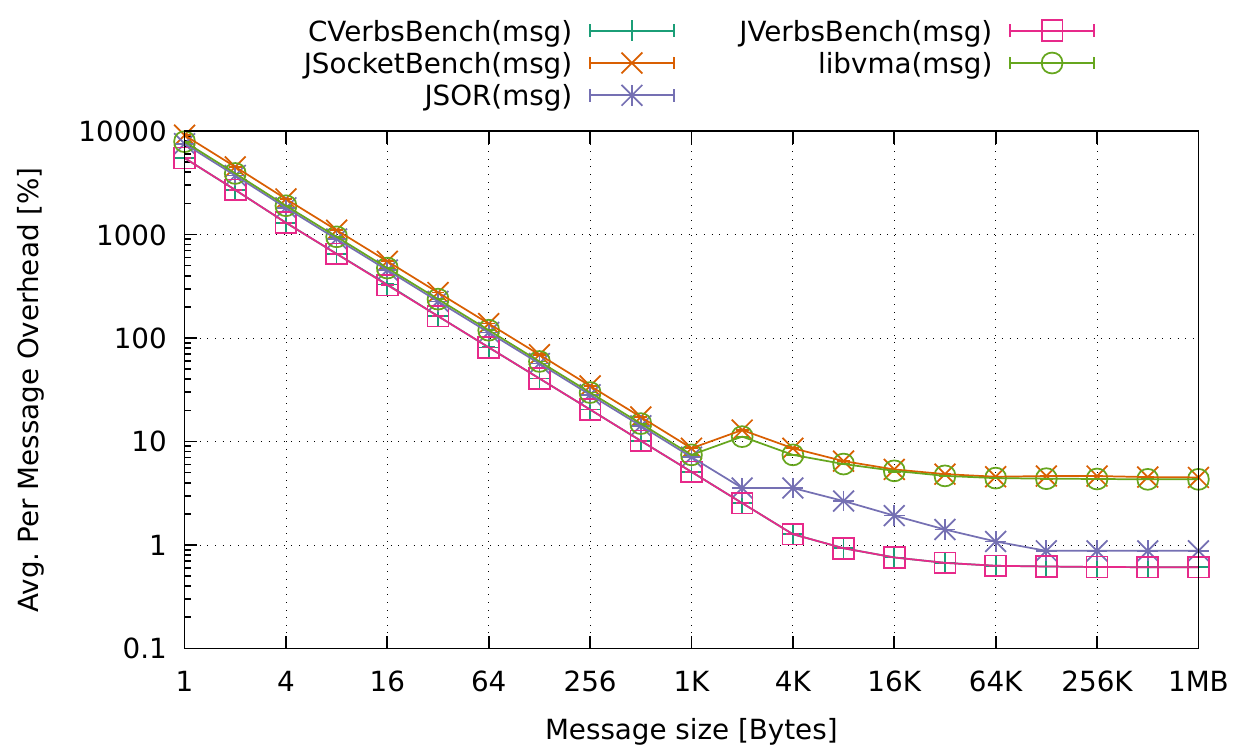}
	\caption{Average per message overhead in percentage in relation to the payload size transferred.}
	\label{fig_eval_overhead_verb}
\end{figure}

IPoIB and libvma are implementing buffer/message aggregation when sending data. Applications on high load sending many small messages benefit highly from increased throughput and the overall per message overhead is lowered. However, in order to determine the general per message overhead, we used the pingpong benchmark which does not allow aggregation due to its nature. The results of both types (sockets/verbs) are depicted in a single figure (see Figure \ref{fig_eval_overhead_verb}).

We try to give a rough breakdown of the overhead involved with each method evaluated. A precise breakdown is rather difficult with just the raw amount of data captured from the ports as re-transmission of packages are also captured (e.g. RC queue pairs or custom protocols based on UD queue pairs).

The results in Figure \ref{fig_eval_overhead_verb} show that the overhead for msg operations of C-verbs and jVerbs are identical. For a single byte of payload, an additional 54 bytes are required which corresponds to two 27 byte headers which are part of the low-level InfiniBand protocol. Required by the RC protocol, one package is used for sending the ping and the other package to receive the pong. The metadata consists of a local routing header (8 bytes), base transport header (12 bytes), invariant CRC (4 bytes) and variant CRC (2 bytes) \cite{ibspec}. This makes a total of 26 bytes which is close to the measured 27 bytes (errors due to \textit{port\_xmit\_data} yielding values in octets). For RDMA operations, an additional RDMA extended transport header (16 bytes) is added which makes a total of 42 bytes of ``metadata`` for such a packet. Naturally, this overhead cannot be avoided. As expected with jVerbs using the native verbs directly without adding another software protocol layer, the overhead added is identical to C-verbs's. The overhead stays constant which leads to an overall decreasing per message overhead with increasing payload size. Starting with 8 kB payload size, the overhead ratio drops below 1\%. 

The overhead of the socket-based solutions is overall slightly higher. Again, considering 1 byte messages, JSOR adds an additional ~7500 \%, libvma 7900 \% and IPoIB 9100 \% overhead to each pingpong transfer. libvma and IPoIB rely on UD messaging verbs which add a datagram extended transport header (8 bytes) to the InfiniBand header and include additional information to allow IP-address based routing of the packages. The IPoIB specification describes an additional header of 4 octets (4 bytes) and IP header (e.g. IPv4 20 bytes + 40 bytes optional) which are added alongside the message payload \cite{ipoib}. libvma adds an IP-address (4 bytes) and Ethernet frame header (14 bytes) \cite{libvma}. Remaining data is likely committed towards a software signalling protocol. Regarding JSOR, we could not find any information about the protocol implemented (closed source).

\subsection{Uni-directional Throughput}
\label{eval_uni}

\begin{figure}
	\centering
	\includegraphics[width=3.3in]{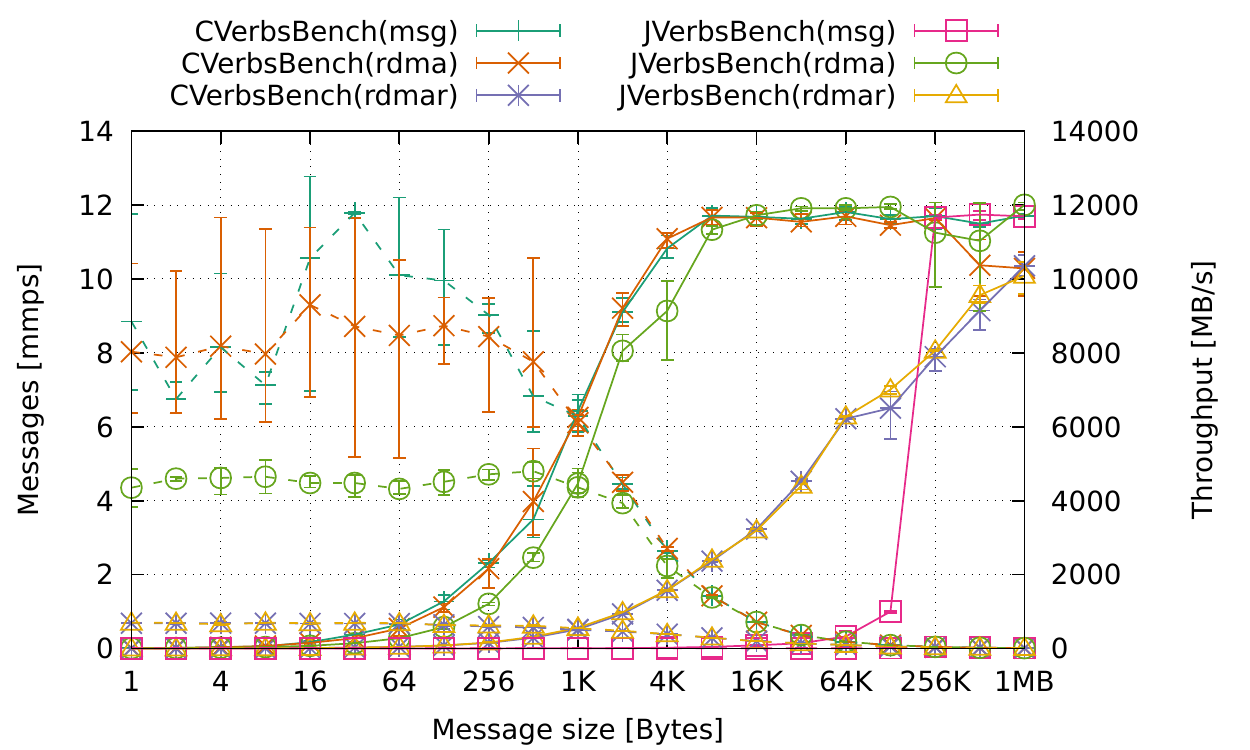}
	\caption{\textbf{Uni-directional} throughput, \textbf{verbs-based} libraries with different transfer methods, increasing message size, \textbf{100 Gbit/s}.}
	\label{fig_eval_verb_tp_uni}
\end{figure}


This section presents the throughput results of the uni-directional benchmark. Starting with the verbs-based results depicted in Figure \ref{fig_eval_verb_tp_uni}, jVerbs RDMA write message throughput (4.3 - 4.6 mmps) is about half of C-verbs's RDMA write throughput (7.9 - 9.3 mmps) for small payload sizes up to 512 byte. The RDMA write performance of C-verbs is nearly double the throughput of C-verbs messaging but with high jitter. Starting at 1 kB payload size, jVerbs's RDMA write throughput stays clearly below both C-verbs's RDMA write and message send throughput. Interesting to note that C-verbs's messaging is significantly better, though highly jittery, on small messages up to 512 bytes and middle sized messages up to 4 kB. Both C-verbs operations saturate throughput with 8 kB payload size and low jitter at around 11.7 GB/s. We could not determine the reason for the very poor performance of jVerbs's msg verbs on both 56 Gbit/s and 100 Gbit/s hardware.

\begin{figure}
	\centering
	\includegraphics[width=3.3in]{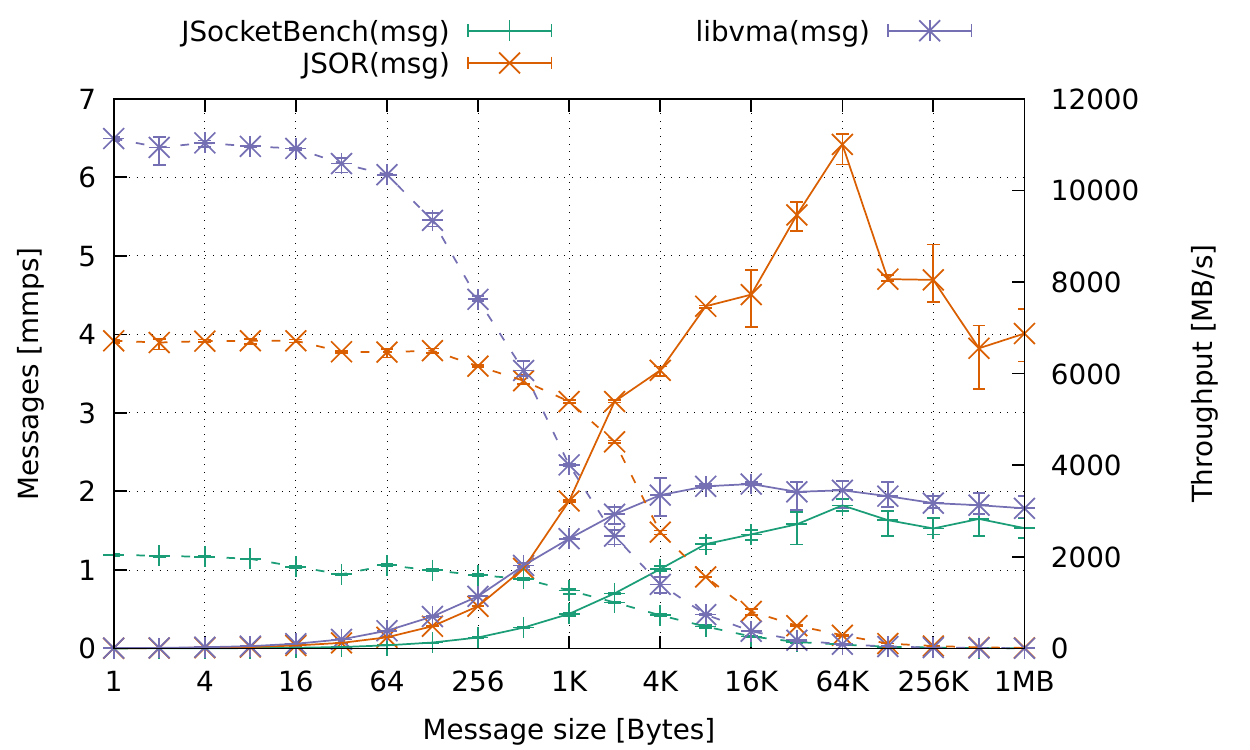}
	\caption{\textbf{Uni-directional} throughput, \textbf{socket-based} libraries, increasing message size, \textbf{100 Gbit/s}.}
	\label{fig_eval_sock_tp_uni}
\end{figure}


The results of socket-based libraries are depicted in Figure \ref{fig_eval_sock_tp_uni}. On small payload sizes up to 256 bytes, IPoIB achieves a throughput of approx. 1 - 1.2 mmps. With increasing payload size, the throughput starts saturating at 32 kB message size and peaks at 64 kB with 3.1 GB/s throughput. The results of libvma show an highly increased throughput of 6.0 to 6.5 mmps for up to 64 byte messages. Overall throughput for middle and large sized messages surpasses IPoIB's peaking at 3.5 GB/s with 8 kB messages but also starting to decrease down to 3.0 GB/s when increasing the message size up to 1 MB. JSOR achieves a significantly lower throughput of 3.8 - 3.9 mmps for up to 128 byte messages. However, JSOR provides a much higher throughput starting at 1 kB message size compared to IPoIB and libvma. Throughput peaks at 64 kB message size with 11 GB/s but drops down to approx 6.5 GB/s with 512 kB messages afterwards. As described in Section \ref{evaluation}, we determined that JSOR's performance degrades considerable on payload sizes of 128 kB and greater which required us to increase the RDMA buffer size (to 1 MB). However, this problem could not be resolved entirely. 

The results on 56 Gbit/s hardware for both, verbs and sockets, show an overall and expected lower throughput but not any notable differences. Thus and due to space constraints, the figures are omitted. But, results for libvma were not available because the benchmarks failed repeatedly with a non fixable ''connection reset`` error by libvma.

\subsection{Bi-directional Throughput}
\label{eval_bi}

\begin{figure}
	\centering
	\includegraphics[width=3.3in]{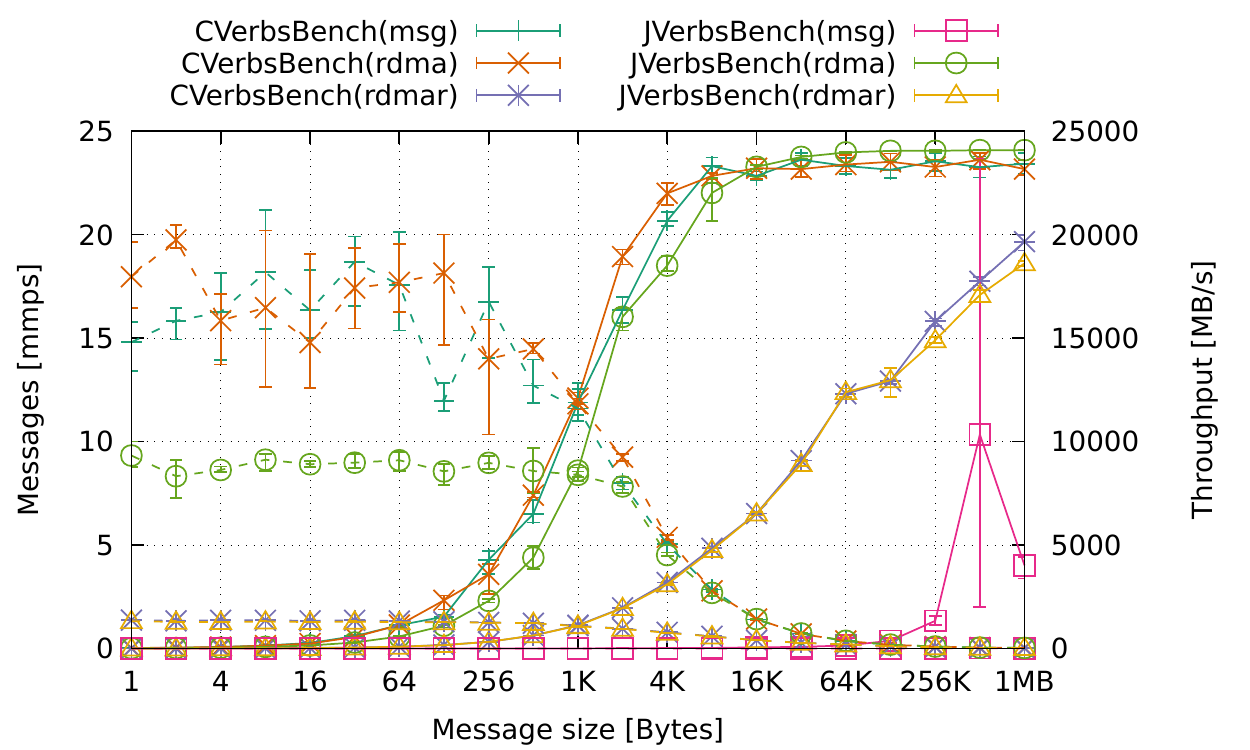}
	\caption{\textbf{Bi-directional} throughput, \textbf{verbs-based} libraries with different transfer methods, increasing message size, \textbf{100 Gbit/s}.}
	\label{fig_eval_verb_tp_bi}
\end{figure}

\begin{figure}
	\centering
	\includegraphics[width=3.3in]{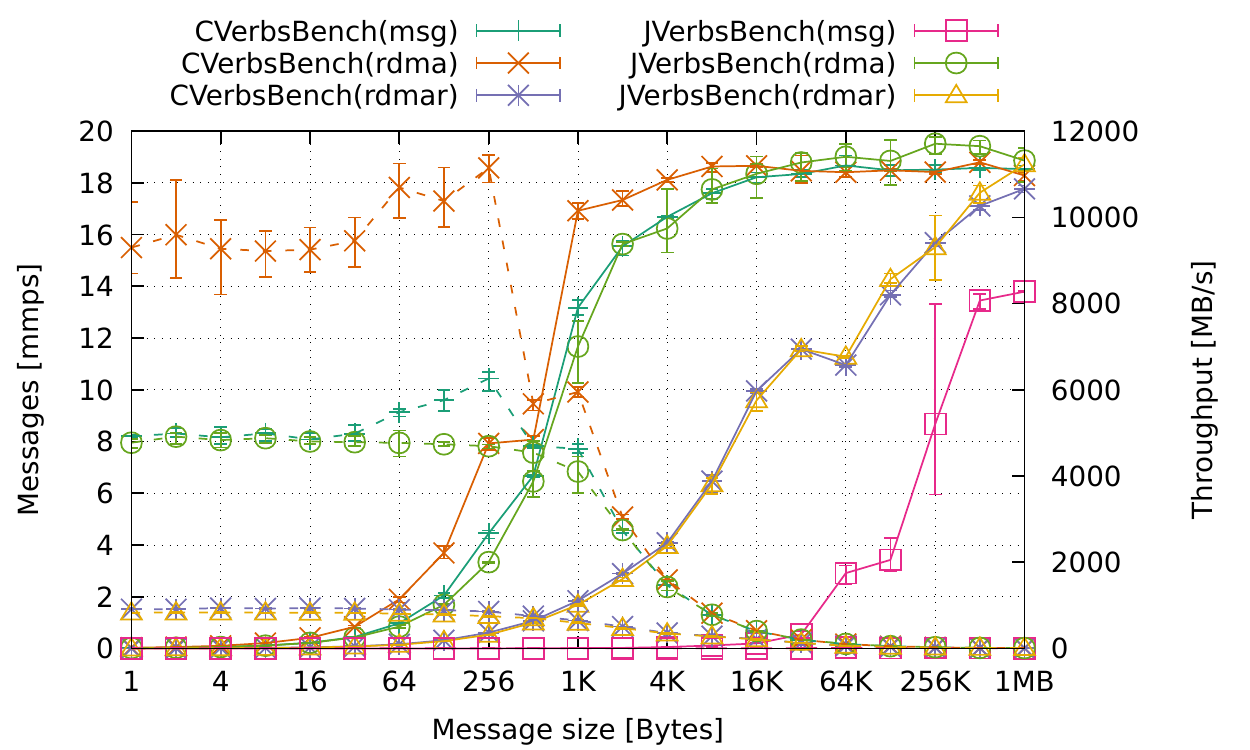}
	\caption{\textbf{Bi-directional} throughput, \textbf{verbs-based} libraries with different transfer methods, increasing message size, \textbf{56 Gbit/s}.}
	\label{fig_eval_verb_tp_bi_56}
\end{figure}

This section presents the throughput results of the bi-directional benchmark. With full-duplex communication supported, we expect roughly double the throughput of the uni-directional results in general. Figure \ref{fig_eval_verb_tp_bi} depicts the results of the verbs-based implementations and, as expected, all implementations show roughly double the message rate on small messages and roughly double the throughput on large messages compared to the uni-directional results (\S \ref{eval_uni}). 

C-verbs RDMA writes are still jittery but yield the best performance with 15.8 - 19.7 mmps for 1 - 128 byte payload size, 23.1 GB/s peak performance at 32 kB payload size. This is followed by C-verbs messages with 11.5 - 18.6 mmps for 1 - 512 bytes, 23.3 GB/s peak performance at 8 kB payload size. jVerbs RDMA writes perform worse on payload sizes up to 1 kB (8.3 - 9.3 mmps) but with less jitter than C-verbs RDMA writes. Saturation on large messages starts at around around 16 kB with 23.7 GB/s with a peak performance of 24.0 GB/s at 128 kB message size. RDMA reads of both verbs interfaces are nearly on par. The incomprehensible poor performance of jVerbs msg verbs, as already seen on the uni-directional benchmark results (\S \ref{eval_uni}), is also present here.

The 56 Gbit/s results are depicted in Figure \ref{fig_eval_verb_tp_bi_56} and show an overall similar but as expected lower performance regarding throughput. On small messages, the RDMA write performance of C-verbs is less jittery and jVerb's not significantly lower compared to the 100 Gbit/s results. The RDMA write performance of C-verbs clearly outperforms jVerb's sometimes slight jittery performance on 128 byte to 1 kB messages.

\begin{figure}
	\centering
	\includegraphics[width=3.3in]{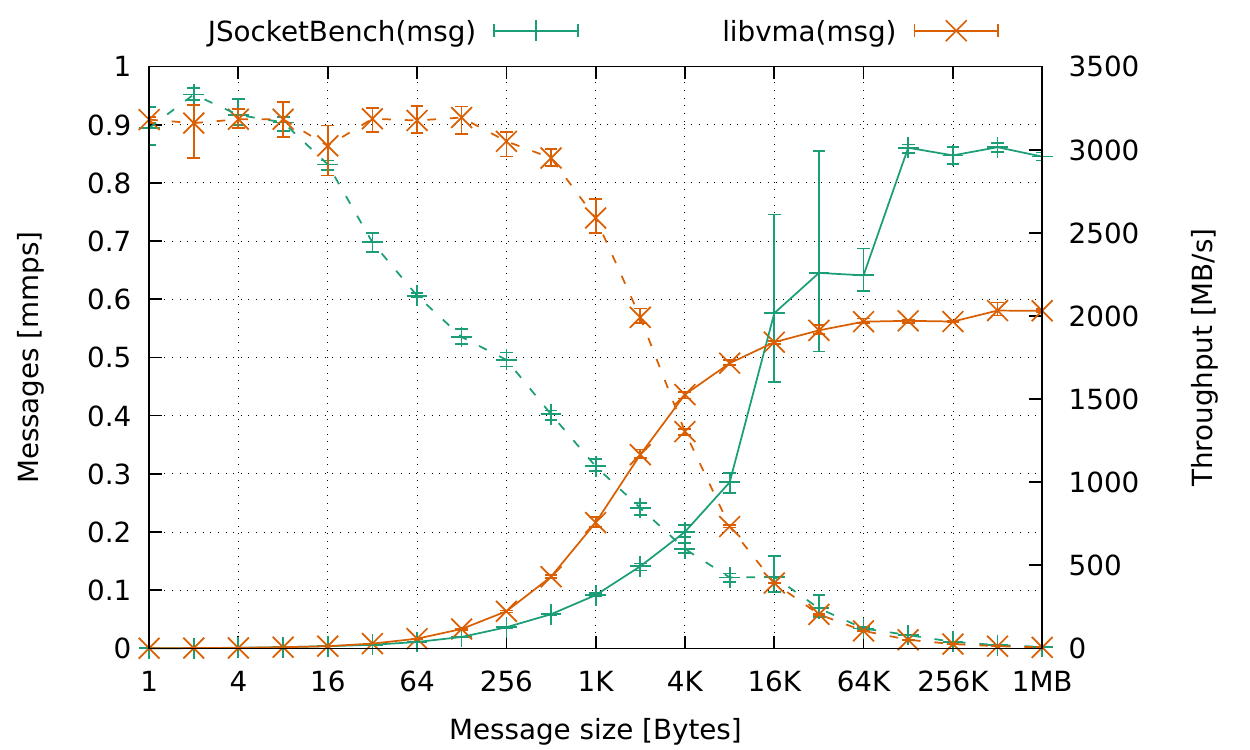}
	\caption{\textbf{Bi-directional} throughput, \textbf{socket-based} libraries, increasing message size, \textbf{100 Gbit/s}.}
	\label{fig_eval_sock_tp_bi}
\end{figure}

\begin{figure}
	\centering
	\includegraphics[width=3.3in]{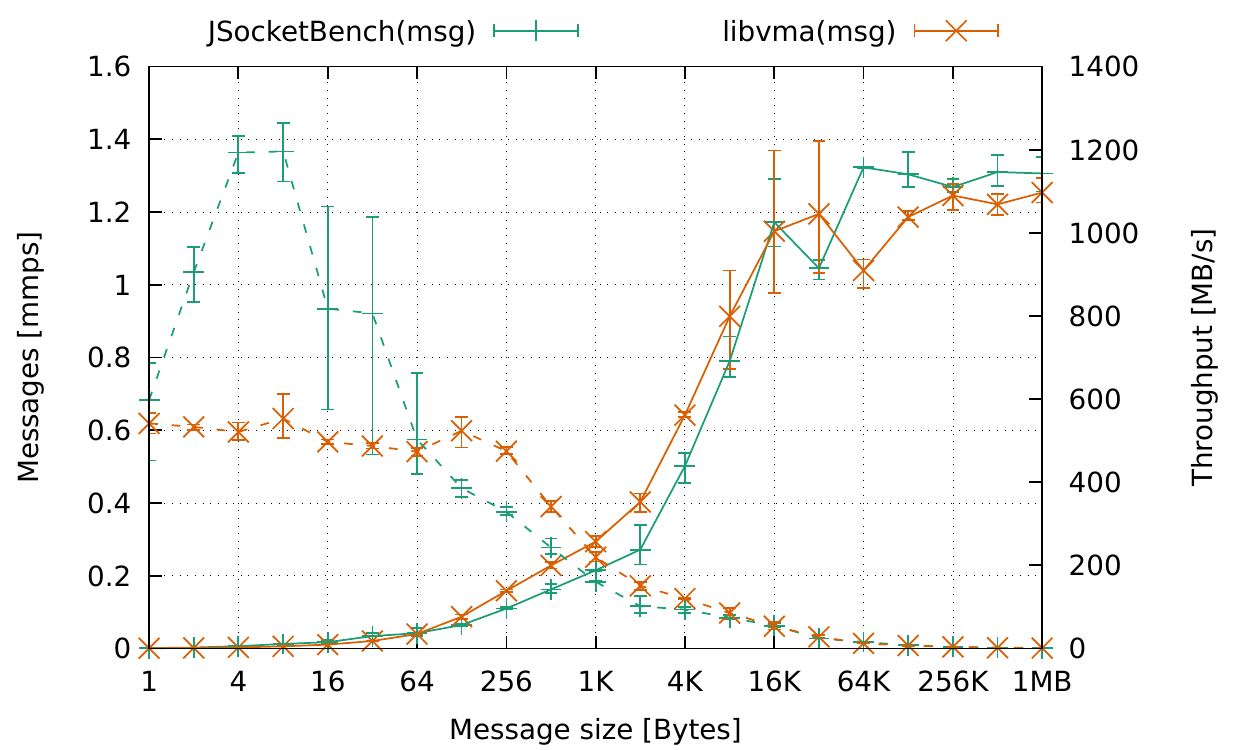}
	\caption{\textbf{Bi-directional} throughput, \textbf{socket-based} libraries, increasing message size, \textbf{56 Gbit/s}.}
	\label{fig_eval_sock_tp_bi_56}
\end{figure}

Figure \ref{fig_eval_sock_tp_bi} depicts the socket-based results of the bi-directional benchmark. Due to unresolvable errors causing disconnects, especially on message sizes smaller 512 bytes, we cannot provide results for JSOR. This seems to be a known problem \cite{jsorissue1} but increasing the send and receive queue sizes as described does not resolve this issue. Furthermore, the benchmark does not terminate anymore on message sizes greater than 32 kB. The proposed solution to increase the buffer size does not resolve this problem either \cite{jsorissue2}.

The results available show a very low overall performance for IPoIB and libvma compared to their results on the uni-directional benchmark (\S \ref{eval_uni}). IPoIB achieves an aggregated throughput of 0.89 to 0.95 mmps for just up to 16 byte messages with a considerable drop in performance for small messages afterwards. But, throughput increases with increasing message size starting with middle sized messages saturating and peaking at 128 kB message with 3.0 GB/s aggregated throughput. Further notable are high fluctuations for 16 kB to 64 kB. On small messages, libvma can at least provide a constant performance for small messages up to 128 bytes with 0.9 mmps. Throughput increases with increasing message size with saturation starting at approx. 32 kB messages with 1.9 GB/s aggregated throughput. A lower peak performance than IPoIB is reached at 512 kB messages with 2.0 GB/s.

The results on 56 Gbit/s hardware in Figure \ref{fig_eval_sock_tp_bi_56} show high fluctuations on up to 64 byte messages with IPoIB and also on 8 kB to 32 kB messages with libvma.

\subsection{One-sided Latency}
\label{eval_lat}

\begin{figure}
	\centering
	\includegraphics[width=3.3in]{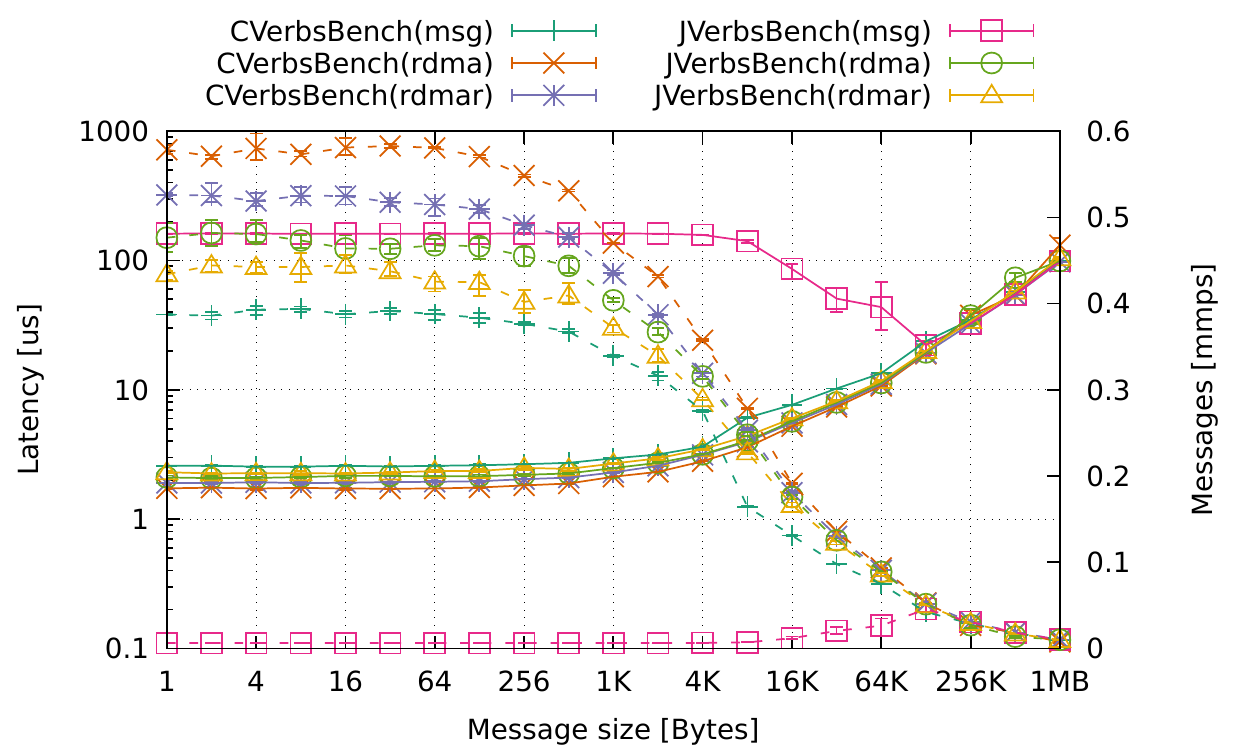}
	\caption{Average \textbf{one-sided latency}, \textbf{verbs-based} libraries with different transfer methods, increasing message size, \textbf{100 Gbit/s}.}
    \label{fig_eval_verb_lat_avg}
\end{figure}

\begin{figure}
	\centering
	\includegraphics[width=3.3in]{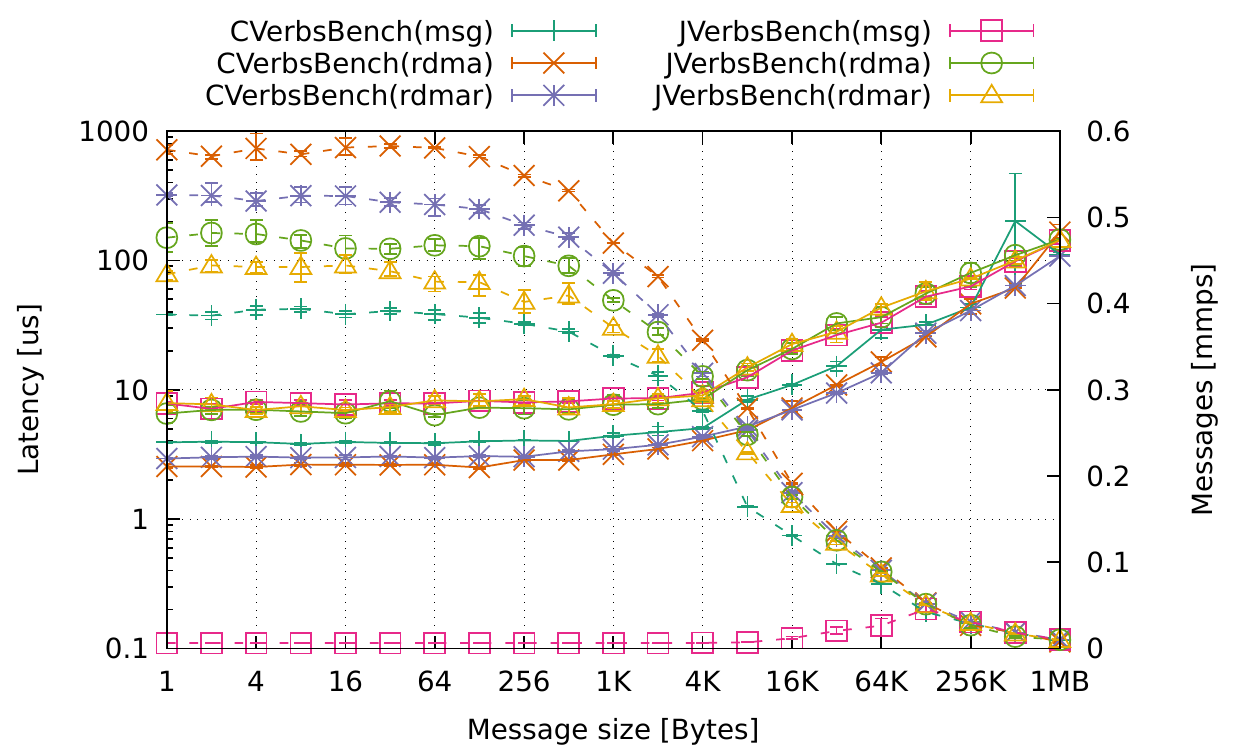}
	\caption{99.9th percentile \textbf{one-sided latency}, \textbf{verbs-based} libraries with different transfer methods, increasing message size, \textbf{100 Gbit/s}.}
	\label{fig_eval_verb_lat_999}
\end{figure}

\begin{figure}
	\centering
	\includegraphics[width=3.3in]{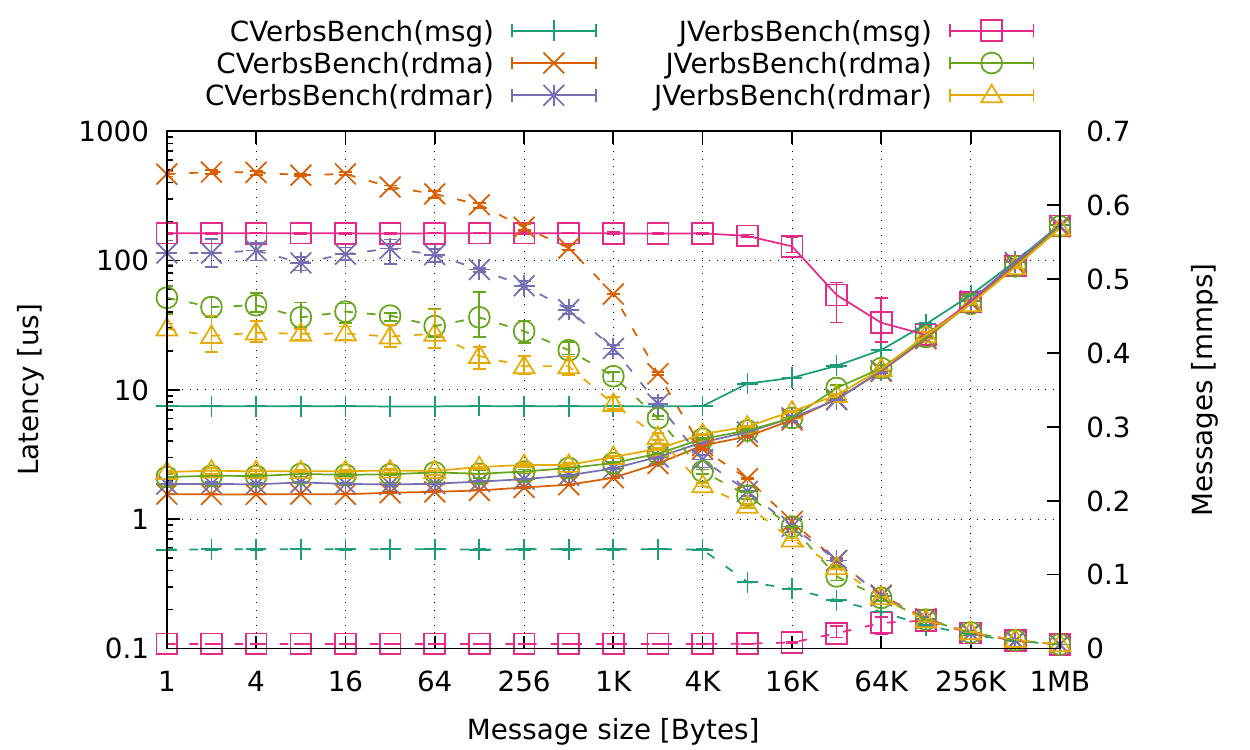}
	\caption{Average \textbf{one-sided latency}, \textbf{verbs-based} libraries with different transfer methods, increasing message size, \textbf{56 Gbit/s}.}
	\label{fig_eval_verb_lat_avg_56}
\end{figure}


This section presents the results of the one-sided latency benchmark to determine the latency of a single operation. Section \ref{eval_pingpong} further discusses full RTT latency for a ping-pong communication pattern. Results are separated by socket-based and verbs-based, and include the average latency as well as the 99.9th percentiles.

Figure \ref{fig_eval_verb_lat_avg} shows the average latencies and Figure \ref{fig_eval_verb_lat_999} the 99.9th percentiles of the verbs-based benchmarks. The results of all native verbs-based communication as well as jVerbs RDMA write and read are as expected providing an average close to 1 \textmu s latency. From lowest to highest: C-verbs RDMA write, C-verbs RDMA read, jVerbs RDMA write, jVerbs RDMA read and C-verbs msg. As expected, jVerbs adds some overhead leading to a minor increase in latency. 

But, jVerbs msg shows unexpected average latency results. Up to 4 kB message size, which equals the used MTU size, the latency is very high and constant at approx. 160 \textmu s. With further increasing message size, the latency is lowered and approximates the average latencies of the other transfer methods. At 128 kB message size, it goes along with the other results. This is also present on the results on 56 Gbit/s hardware for jVerbs msg (see Figure \ref{fig_eval_verb_lat_avg_56}).

\begin{figure}
	\centering
	\includegraphics[width=3.3in]{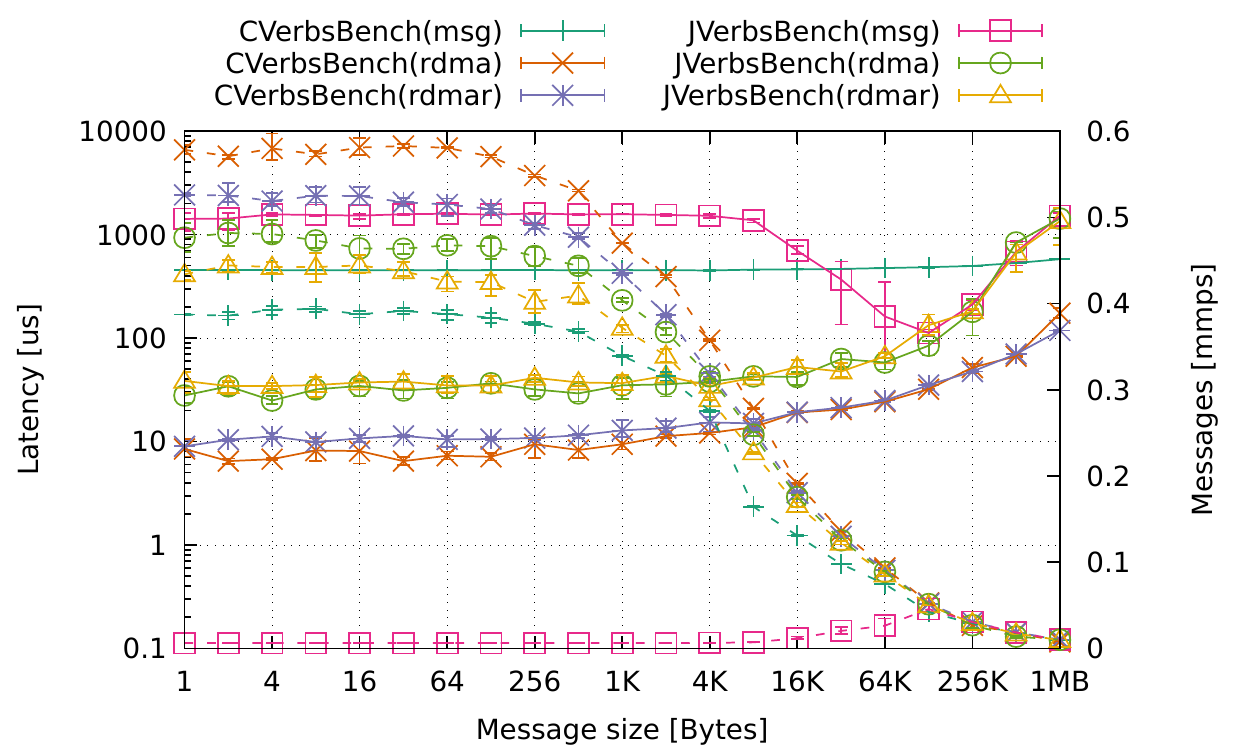}
	\caption{99.99th percentile \textbf{one-sided latency}, \textbf{verbs-based} libraries with different transfer methods, increasing message size, \textbf{100 Gbit/s}.}
	\label{fig_eval_verb_lat_9999}
\end{figure}

\begin{figure}
	\centering
	\includegraphics[width=3.3in]{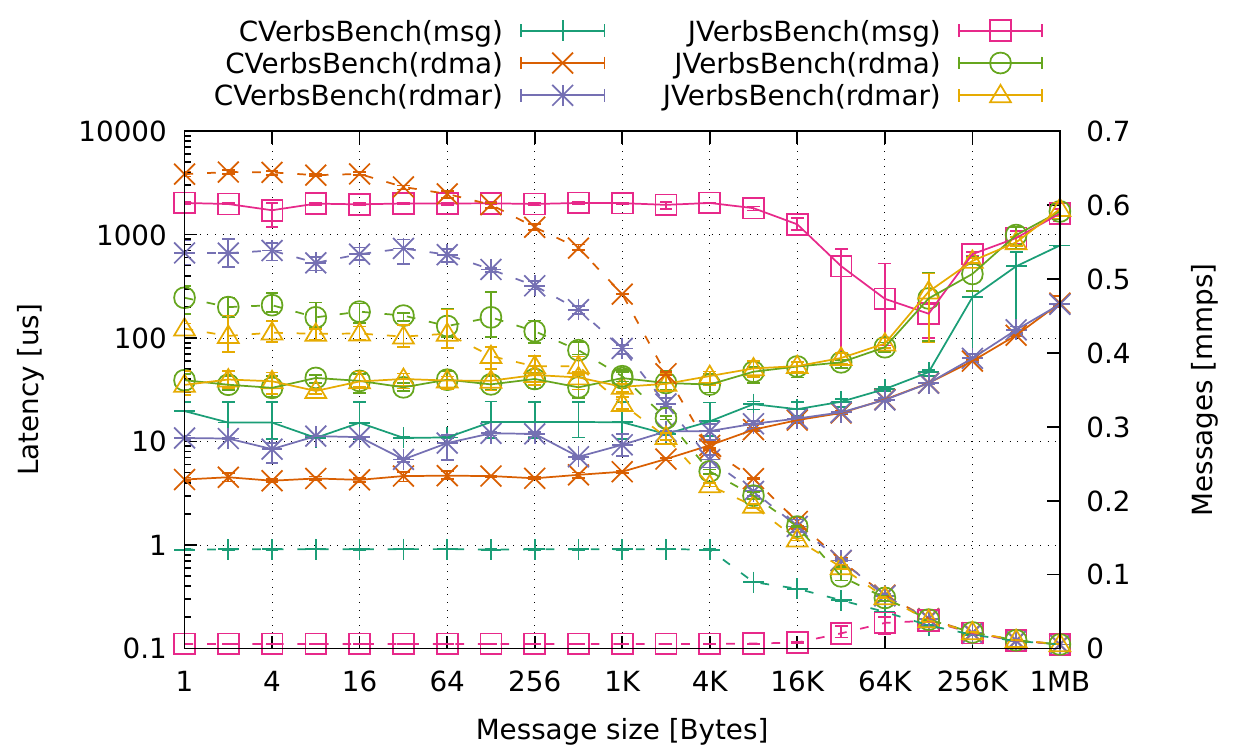}
	\caption{99.99th percentile \textbf{one-sided latency}, \textbf{verbs-based} libraries with different transfer methods, increasing message size, \textbf{56 Gbit/s}.}
	\label{fig_eval_verb_lat_9999_56}
\end{figure}

To further analyze this issue, we depicted the 99.9th percentiles in Figure \ref{fig_eval_verb_lat_999}. The other transfer methods are showing expected behaviour and overall low latency. But, jVerbs msg also shows very low latencies around 8 \textmu s latency for 99.9th of all messages transferred. Thus, just a small amount of messages yields latencies higher than 8 \textmu s. This can be verified by analyzing the 99.99th percentiles (i.e. 1,000 worst out of 10 million) depicted in Figure \ref{fig_eval_verb_lat_9999}. The results show a latency of approx. 1500 \textmu s confirming that there is a very small amount of messages with very high latency causing a rather overall high average latency. This conclusion can also be drawn for jVerbs's results on 56 Gbit/s hardware (see Figure \ref{fig_eval_verb_lat_9999_56}).

Two further issues regarding C-verbs msg can be observed. The first is a nearly constant 450 \textmu s on the 99.99th percentile results on 100 Gbit/s hardware (see Figure \ref{fig_eval_verb_lat_9999}). But, the average and 99.9th percentile results of C-verbs msg are as expected. This means that 1,000 out of 10,000,000 messages show a latency of 450 \textmu s or worse. This is caused by the RC protocol yielding occasional RNR NAKs on high loads when sending data and no corresponding work requests for receiving are queued on the remote at that moment. This is proven by the value of the hardware counter \textit{rnr\_nak\_retry\_err} which shows that about 3,100 send requests are NAK'd during one benchmark run. For the sender, each NAK enforces a small delay to wait for resources to become available again. This behaviour cannot be avoided in such a benchmark scenario. 

The second issue with C-verbs msg is a rather high average latency of 7.5 \textmu s for up to 4 kB messages on 56 Gbit/s hardware. However, the RDMA write/read latency results on 56 Gbit/s hardware are similar to the ones on 100 Gbit/s hardware. The 99.9th and 99.99th percentiles (10 to 15 \textmu s) prove that the base latency on that hardware configuration with 56 Gbit/s speed is unexpectedly high.

\begin{figure}
	\centering
	\includegraphics[width=3.3in]{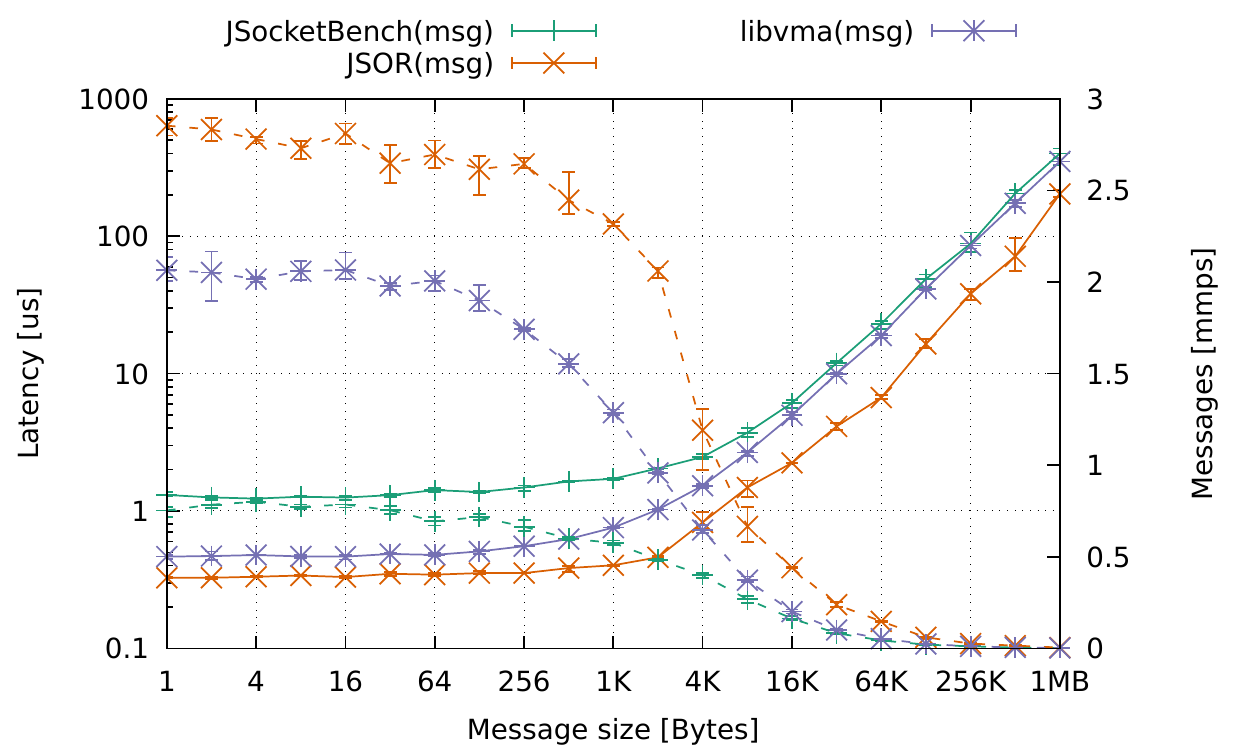}
	\caption{Average \textbf{one-sided latency}, \textbf{socket-based} libraries, increasing message size, \textbf{100 Gbit/s}.}
	\label{fig_eval_socket_lat_avg}
\end{figure}

\begin{figure}
	\centering
	\includegraphics[width=3.3in]{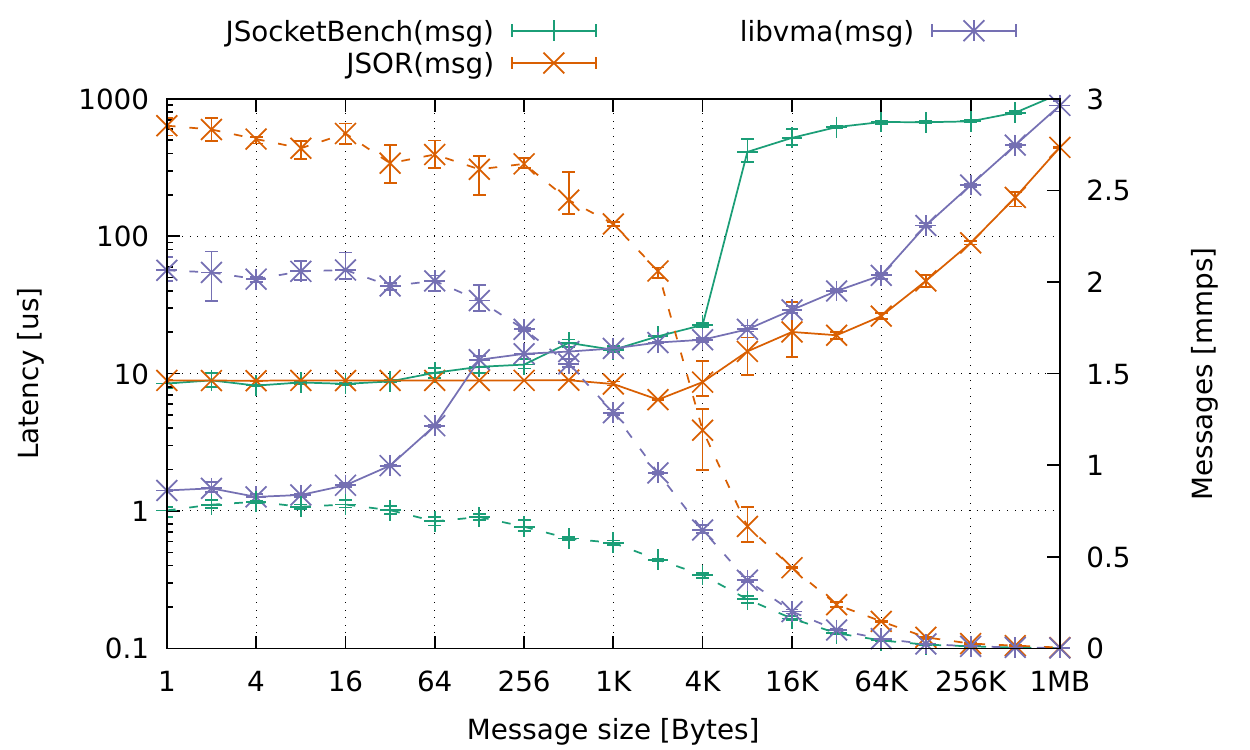}
	\caption{99.9th percentile \textbf{one-sided latency}, \textbf{socket-based} libraries, increasing message size, \textbf{100 Gbit/s}.}
	\label{fig_eval_socket_lat_999}
\end{figure}

\begin{figure}
	\centering
	\includegraphics[width=3.3in]{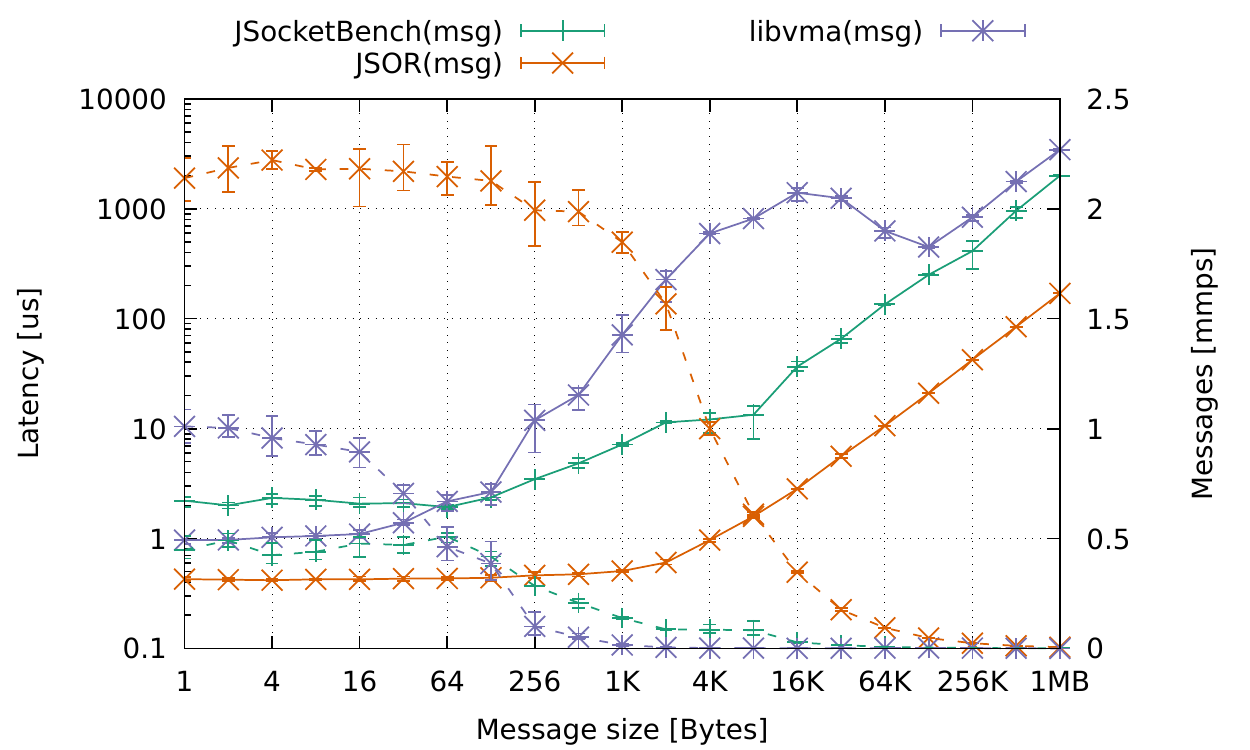}
	\caption{Average \textbf{one-sided latency}, \textbf{socket-based} libraries, increasing message size, \textbf{56 Gbit/s}.}
	\label{fig_eval_socket_lat_avg_56}
\end{figure}


For the socket-based solutions, the average latencies in Figure \ref{fig_eval_socket_lat_avg} show that JSOR performs best with an average per operation latency of 0.3 - 0.4 \textmu s for up to 1 kB messages. With further increasing payload size, latency increases as expected. libvma shows similar results with a slightly higher latency of 0.4 - 0.5 \textmu s for small messages. IPoIB follows with a further increased average latency of 1.3 to 1.5 \textmu s for small messages. These results, especially JSOR's, seem unexpected low at first glance. But, when considering the socket-interface, it does not provide means to return any feedback to the application when data is actually sent to the remote. With verbs, one polls the completion queue and as soon as the work completion is received, it is guaranteed that the local data is sent and received by the remote. A socket send-call however, does not guarantee that the data is sent once it returns control to the caller. Typically, a buffer is used to allow aggregation of data before putting it on the wire. JSOR, libvma and IPoIB implement message aggregation before actually sending out any data. This is further proven by the ping-pong benchmark which does not allow any aggregation to be applied by the backend (\S \ref{eval_pingpong}). On 56 Gbit/s hardware (see Figure \ref{fig_eval_socket_lat_avg_56}), JSOR and IPoIB show similar results with a slightly higher latency but libvma shows significantly worse results for 256 byte to 64 kB message sizes compared to running on 100 Gbit/s hardware.

The 99.9th percentiles in Figure \ref{fig_eval_socket_lat_999} show further interesting aspects not just limited to message aggregation. libvma starts with very low 99.9th latencies for up to 16 byte messages indicating a rather low threshold for aggregation benefitting small messages by keeping the total of worst latencies low. However, with 16 to 128 byte messages, the latency increases considerably. JSOR and IPoIB show similar results for small messages up to 1 kB. JSOR's stays even constant with 9 \textmu s up to 512 byte messages. This indicates a flush threshold based on the number of messages instead of a total buffer size. But, the latency of IPoIB starts to increase already starting with 64 byte message size and even jumps significantly up to over 400 \textmu s with 8 kB message size. This might indicate that additional allocations are involved for large(r) messages increasing the overall worst latencies significantly. On 56 Gbit/s hardware, JSOR's and IPoIB's results are similar with a slightly higher latency. But, libvma's 99.9th percentile latency is significantly increasing with 256 byte messages and even reaching 600 ms on multiple sizes greater than 4 kB.

\subsection{Ping-pong Latency}
\label{eval_pingpong}

\begin{figure}
	\centering
	\includegraphics[width=3.3in]{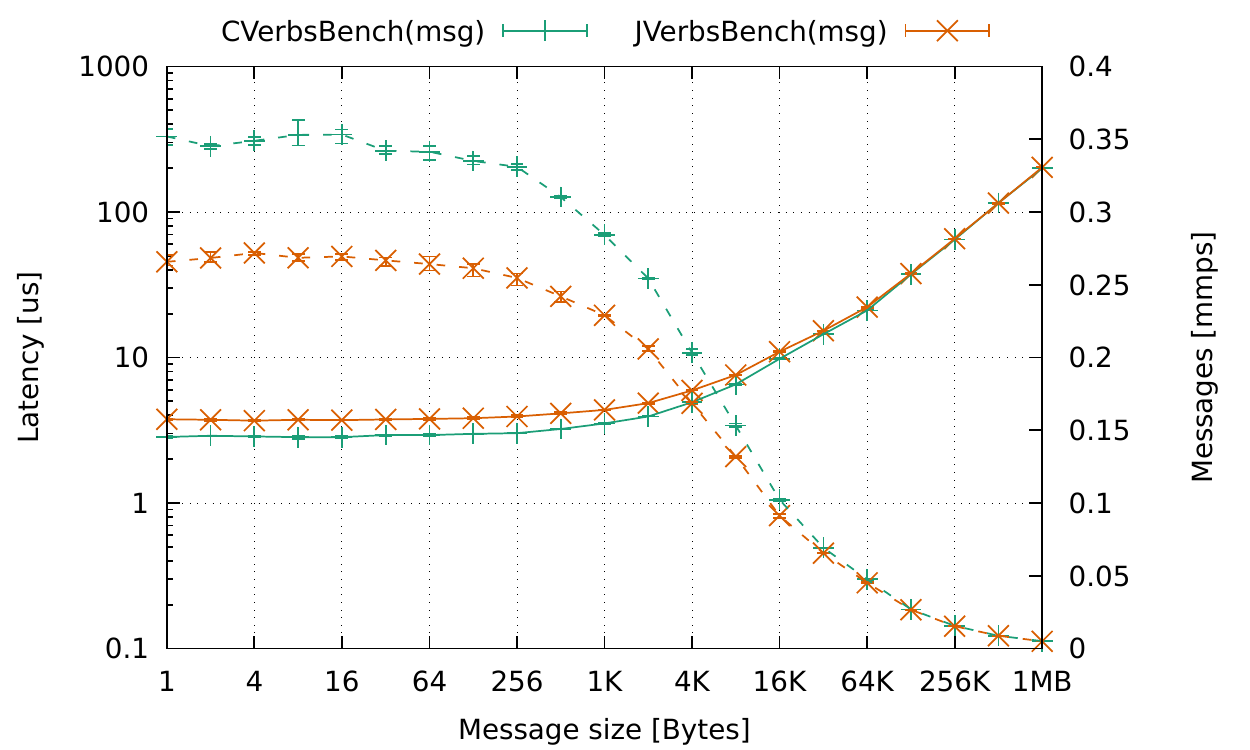}
	\caption{Average \textbf{ping-pong} latency, \textbf{verbs-based} libraries with different transfer methods,  increasing message size, \textbf{100 Gbit/s}.}
	\label{fig_eval_verbs_pp_avg}
\end{figure}


\begin{figure}
	\centering
	\includegraphics[width=3.3in]{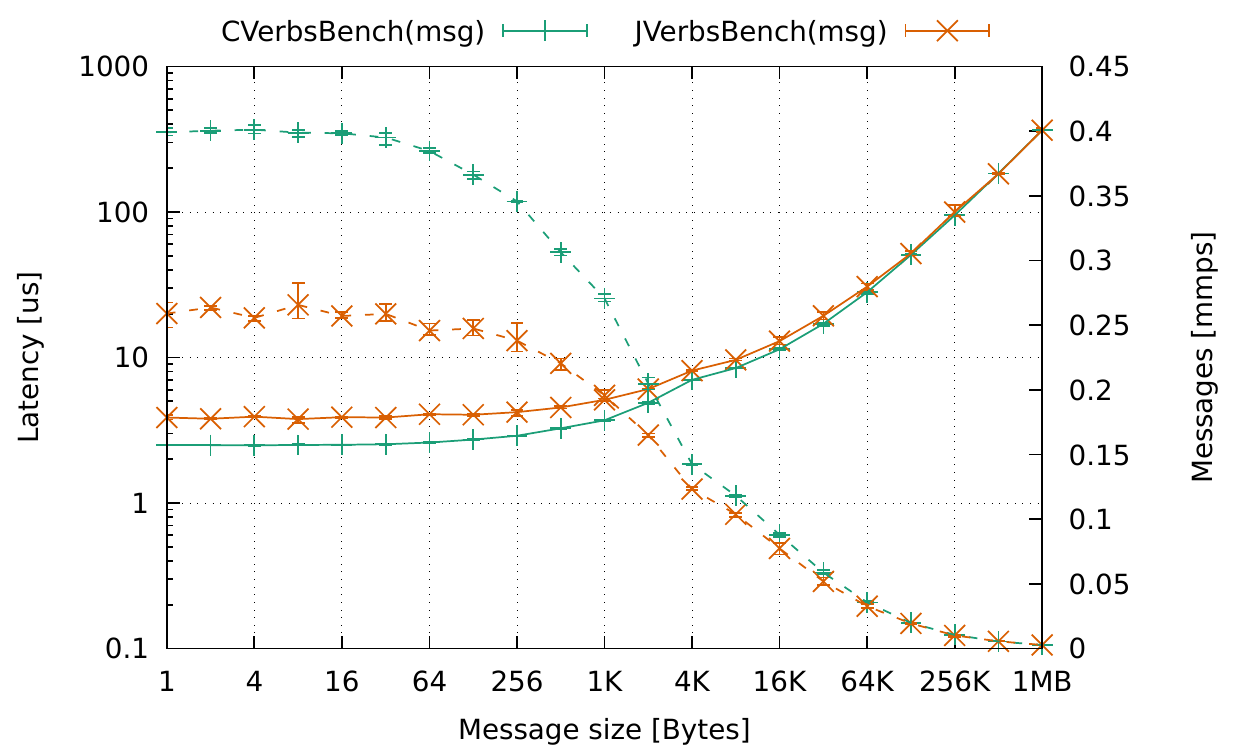}
	\caption{Average latency \textbf{ping-pong}, \textbf{verbs-based} libraries with different transfer methods, increasing message size, \textbf{56 Gbit/s}.}
	\label{fig_eval_verbs_pp_avg_56}
\end{figure}


In this section, we present the results of the ping-pong latency benchmark. Due to the nature of the communication pattern, the methods of transfer are limited to messaging operations for verbs-based implementations. Using RDMA operations is also possible but requires additional data structures and control logic which is currently not implemented. The average latencies, i.e. full round-trip-times, are depicted in Figure \ref{fig_eval_verbs_pp_avg} (Figure \ref{fig_eval_verbs_pp_avg_56} for 56 Gbit/s results). C-verbs messaging achieves an average latency of 2.8 to 3.2 \textmu s for up to 512 byte messages. jVerbs shows similar behaviour but with a slightly higher latency of 3.7 to 4.1 \textmu s. Further increasing the message size of both, C-verbs and jVerbs increases the latency as expected. The 99.9th percentiles results on 56 Gbit/s and 100 Gbit/s hardware do not show any abnormalities and their Figures are omitted due to space constraints.

\begin{figure}
	\centering
	\includegraphics[width=3.3in]{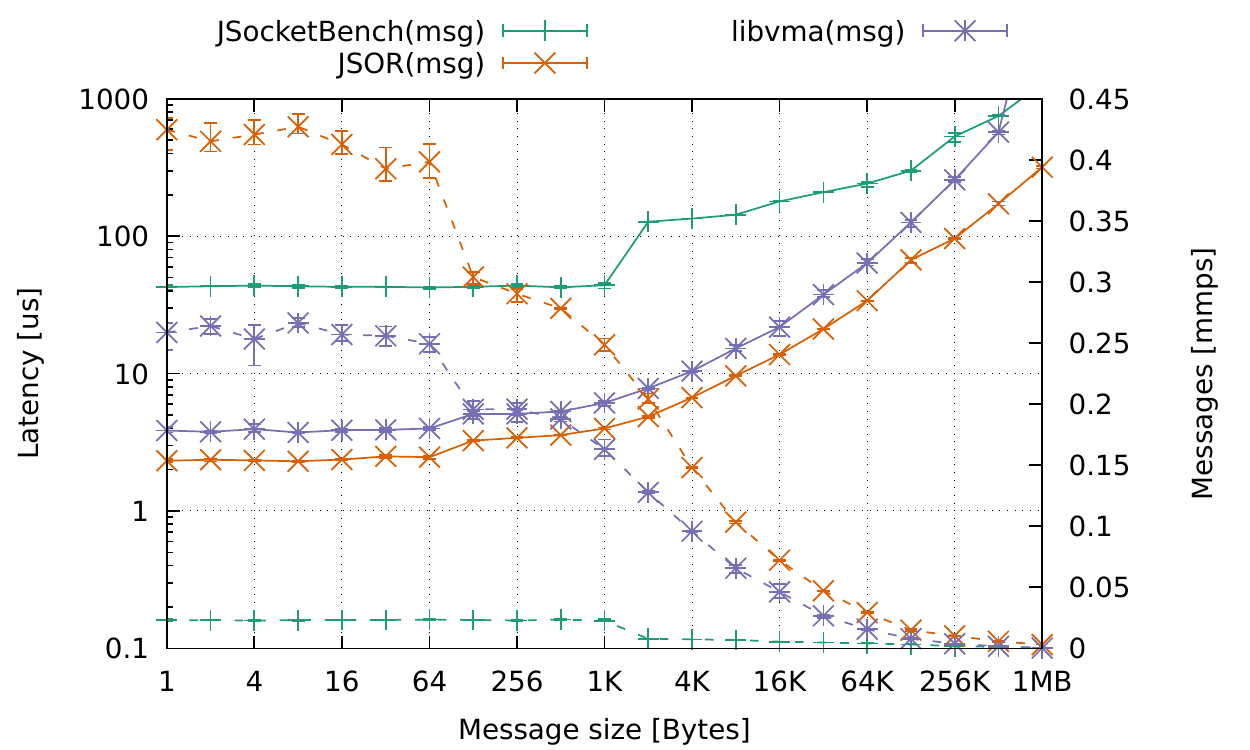}
	\caption{Average \textbf{ping-pong} latency, \textbf{socket-based} libraries, increasing message size, \textbf{100 Gbit/s}.}
	\label{fig_eval_sockets_pp_avg}
\end{figure}

\begin{figure}
	\centering
	\includegraphics[width=3.3in]{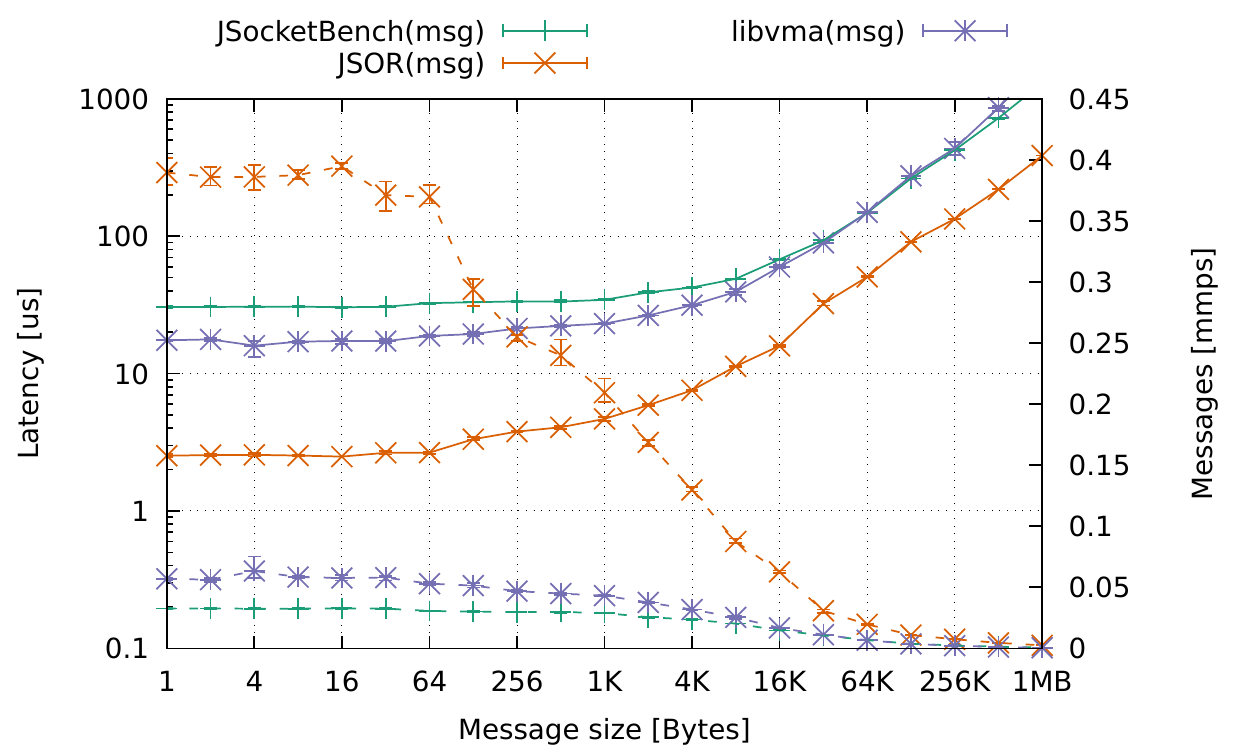}
	\caption{Average \textbf{ping-pong} latency, \textbf{socket-based} libraries, increasing message size, \textbf{56 Gbit/s}.}
	\label{fig_eval_sockets_pp_avg_56}
\end{figure}

The average latencies of the socket-based methods are depicted in Figure \ref{fig_eval_sockets_pp_avg} (Figure \ref{fig_eval_sockets_pp_avg_56} for 56 Gbit/s results). Both, JSOR and libvma show low average latencies of 2.3 to 3.5 \textmu s and 3.7 to 5.3 \textmu s for message sizes up to 512 bytes. With increasing message sizes, the average latency also increases as expected. A small ''latency jump`` of around 1 \textmu s is notable on both libraries from 64 byte to 128 byte message size. IPoIB shows a similar behaviour but with a significant higher average latency of 42.7 to 43.2 \textmu s up to 1 kB message size. The same ''latency jump`` can be seen from 1 kB to 2 kB message size. This increase of latency might be caused due to switching to a different buffer size which might involve additional buffer allocation. The 99.9th percentile results show a similar behaviour without further abnormalities. Their figures are omitted due to space constraints.

%
        
\section{Conclusions}
\label{conclusions}

We presented our JIB-Benchmark to evaluate three socket-based and two verbs-based solutions to leverage InfiniBand in Java applications. We believe that such a benchmark, not available thus far, will help the community and developers interested in using InfiniBand in their Java applications to find a suitable solution for their applications. The benchmark is open source and can be extended to evaluate further libraries. We evaluate the available solutions on two hardware configurations with 56 Gbit/s and 100 Gbit/s InfiniBand NICs. As expected, the socket-based solutions provide a transparent solution requiring low effort to get additional performance from InfiniBand hardware for existing socket-based software without requiring any changes. But, this comes at the price that the full potential of the hardware cannot be exploited, especially on bi-directional communication. Compared to the performance of Gigabit Ethernet, latency is at least halved on 56 Gbit/s hardware and can even be as low as 2-5 \textmu s for small messages. Regarding throughput, one can get an at least ten-fold increase and it is even possible to saturate 56 Gbit/s hardware on uni-directional communication. 

To leverage the true potential of the hardware, the verbs-based solutions are a must. Overall, jVerbs is performing very well and brings nearly native performance on RDMA operations to the Java space with a few minor performance differences. But, the inexplicable limited performance of jVerbs messaging verbs does not allow any meaningful usage in applications. With C-verbs, the full potential of the hardware can be exploited on all communication methods. Thus, one has to decide whether to stay entirely in Java space but having to rely on the proprietary JV9 JVM or having the freedom to write a custom network subsystem using C-verbs with JNI which is more time consuming and challenging.

Our personal recommendations regarding the evaluation: we consider libvma a good solution to benefit from some of the performance of InfiniBand hardware without having to invest a significant amount of time and work and not depending on a proprietary JVM. But, we think that it is worth spending additional work and time on implementing a custom network subsystem based on C-verbs to leverage the true performance of InfiniBand hardware if required for a target application.

\bibliography{paper}
\bibliographystyle{abbrv} 

\end{document}